\documentclass[prb,twocolumn,superscriptaddress]{revtex4-1}
\usepackage{graphicx}
\usepackage{hyperref}
\usepackage{multirow}
\usepackage{verbatim}

\usepackage{fontenc}

\begin{document}

\title{Accelerated materials property predictions and design using motif-based fingerprints}

\author{Tran Doan Huan}
\affiliation{Institute of Materials Science, University of Connecticut, 97 North Eagleville Rd., Unit 3136, Storrs, CT 06269-3136, USA}
\author{Arun Mannodi-Kanakkithodi}
\affiliation{Institute of Materials Science, University of Connecticut, 97 North Eagleville Rd., Unit 3136, Storrs, CT 06269-3136, USA}
\author{Rampi Ramprasad}
\email{rampi@ims.uconn.edu}
\affiliation{Institute of Materials Science, University of Connecticut, 97 North Eagleville Rd., Unit 3136, Storrs, CT 06269-3136, USA}
\date{\today}
\begin{abstract}
Data-driven approaches are particularly useful for computational materials discovery and design as they can be used for rapidly screening over a very large number of materials, thus suggesting lead candidates for further in-depth investigations. A central challenge of such approaches is to develop a numerical representation, often referred to as a fingerprint, of the materials. Inspired by recent developments in chem-informatics, we propose a class of hierarchical motif-based topological fingerprints for materials composed of elements such as C, O, H, N, F, etc., whose coordination preferences are well understood. We show that these fingerprints, when representing either molecules or crystals, may be effectively mapped onto a variety of properties using a similarity-based learning model and hence can be used to predict relevant properties of a material, given that its fingerprint can be defined. Two simple procedures are introduced to demonstrate that the learning model can be inverted to identify the desired fingerprints and then, to reconstruct molecules which possess a set of targeted properties.

\end{abstract}


\maketitle

\section{Introduction}\label{sec:intro}

Data-driven approaches towards materials design and discovery are rapidly increasing in popularity, demand and potency. \cite{Hautier_review, Curtarolo:nature, Vinit_Nature, Rampi:MLbook, hastie:MLbook,Curt:QSPR1, Curt:QSPR2, Tu:QSPR, Curtarolo:03, Rajan200538, schon:ZAAC, Meredig:14, Hansen:13, GuhaBook, Varnek} This emerging trend is fueled by the availability and emergence of large materials databases,\cite{MaterialsProject,Bergerhoff:database,Grazulis2012} as well as our ability to progressively accumulate materials data via high-throughput computations \cite{Ramakrishnan:SD,Pilania_SR} and experiments.\cite{MaterialsProject,Bergerhoff:database,Grazulis2012} Data-driven strategies aimed at rapid property predictions, and ultimately to rational or informed materials design, rely on exploiting the information content of past data, and using such information within heuristic or statistical interpolative learning models to provide estimates of properties of a new material. This approach is entirely analogous to similar pursuits undertaken within chem- and bio-informatics wherein lead candidates worthy of further in-depth investigations are identified rapidly in a first-level of screening.\cite{Rampi:MLbook,hastie:MLbook,GuhaBook}

Data-driven property prediction strategies have two steps. The first involves representing materials numerically via descriptors, attribute vectors, or fingerprints. In the second step, using available ``training" data sets, a mapping is established between the numerical representation of materials and their properties, thus leading to a prediction model. Subsequently, the properties of a new material are estimated using this model after reducing the material to its numerical representation.

One of the central challenges in this whole process is deciding on an appropriate and acceptable numerical representation of materials. The specific choice of this representation is entirely application dependent, and can range from high level descriptors (e.g., $d$-band center, atomic electronegativities) \cite{Andriotis:14,ChiDam} to topological features (e.g., substructural motifs) \cite{Brown:96,Brown:97,Pilania_SR} to microscopic fingerprints that may capture chemical and configurational degrees of freedom (e.g., coulomb matrix, symmetry functions).\cite{CoulombMatrix,Hansen:14,Lilienfeld:RDF,Venke:14} Regardless of the specific choice, the representations are expected to satisfy certain basic requirements. These include invariance of the representation with respect to transformations of the material such as translation, rotation, and permutation of like elements. Moreover, it is desired that the representation be intuitive, elegant and physically and chemically meaningful.

In this contribution, inspired by developments in chem-informatics,\cite{GuhaBook, Varnek} we propose a class of hierarchical motif-based topological fingerprints. This choice, in which the motifs are molecular fragments of varying sizes, is particularly suited to representing molecules and solids composed of elements such as H, C, N, O, F, etc., whose coordination preferences are well understood. Large datasets of molecules and solids are considered, and it is shown that the fingerprints may be effectively mapped to a variety of properties using a similarity based learning algorithm. Moreover, it is demonstrated that the learning model may be inverted to identify fingerprints, and subsequently, to reconstruct actual molecules that possess a desired set of target properties.

\section{Datasets}\label{sec:datasets}
In the present work, we restrict ourselves to systems composed of C, O and H. We used two datasets, one for molecules and one for crystals, to demonstrate the applicability of the proposed fingerprints. Of these two datasets, the former was taken from Ref. \onlinecite{Ramakrishnan:SD} while the latter was prepared by us.

\subsection{Molecule dataset}
A dataset of more than 134,000 small molecules made up of C, O, H, N, and F was reported in Ref. \onlinecite{Ramakrishnan:SD}. This reliable dataset, which contains the optimized geometries, and energetic, electronic, and thermodynamic properties calculated using the B3LYP hybrid exchange-correlation (XC) functional and the 6-31G(2df,p) basis set with the Gaussian 09 software, sets up the stage for many interesting data-mining works.\cite{Lilienfeld:delta,Lilienfeld:molec} A subset of this dataset, containing 45,708 molecules composed of C, O, and H was used in this work. Five properties were considered, including the atomization energy ${\cal E}_{\rm at}$, the energy gap $E_{\rm HL}$ between highest occupied and lowest unoccupied molecular orbitals (HOMO-LUMO gap), the isotropic polarizability $\alpha$, the heat capacity $C_{\rm v}$, and the zero-point vibration energy ${\cal E}_{\rm ZP}$.

\subsection{Crystal dataset}
In addition to the molecules dataset, we prepared another dataset of 215 organic crystals comprising of C, O, and H. This includes
\begin{enumerate}
\item 12 existing polymers composed of C, O, and H,
\item 16 new polymer structures predicted by the minima-hopping method\cite{MHM:OrganovBookChapter,Goedecker:MHM,Amsler:MHM} and USPEX\cite{USPEX} for 16 quasi-one-dimensional polymer chain models reported in Ref. \onlinecite{Vinit_Nature},
\item 34 organic crystals composed of C and H and 153 organic crystals composed of C, O, and H obtained from {\it Crystallography Open Database}. \cite{Grazulis2012}
\end{enumerate}

The obtained structures were optimized by first-principles calculations within the DFT formalism as implemented in Vienna \textit{Ab initio} Simulation Package ({\sc vasp}), \cite{vasp1, vasp2, vasp3, vasp4} utilizing the semi-local rPW86 XC functional \cite{MurrayrPW86} and a plane wave energy cutoff of 400 eV. A Monkhorst-Pack {\bf k}-point mesh\cite{monkhorst} with the spacing of no more than $0.15$\AA$^{-1}$ in the reciprocal space were used for sampling the Brillouin zone, while the van der Waals interactions were estimated with the non-local density functional vdW-DF2. \cite{vdW-DF2} Convergence was assumed when the atomic forces exerting on the atomic sites are smaller than 0.01 eV/\AA. The entire crystals dataset, which includes the optimized structures, the atomization energies ${\cal E}_{\rm at}$, the band gaps $E_{\rm g}$, and the electronic and ionic parts of the dielectric constants, $\epsilon_{\rm elec}$ and $\epsilon_{\rm ion}$, can be found in the Supplemental Material. \cite{supplement}

\begin{figure}[t]
  \begin{center}
    \includegraphics[width= 7 cm]{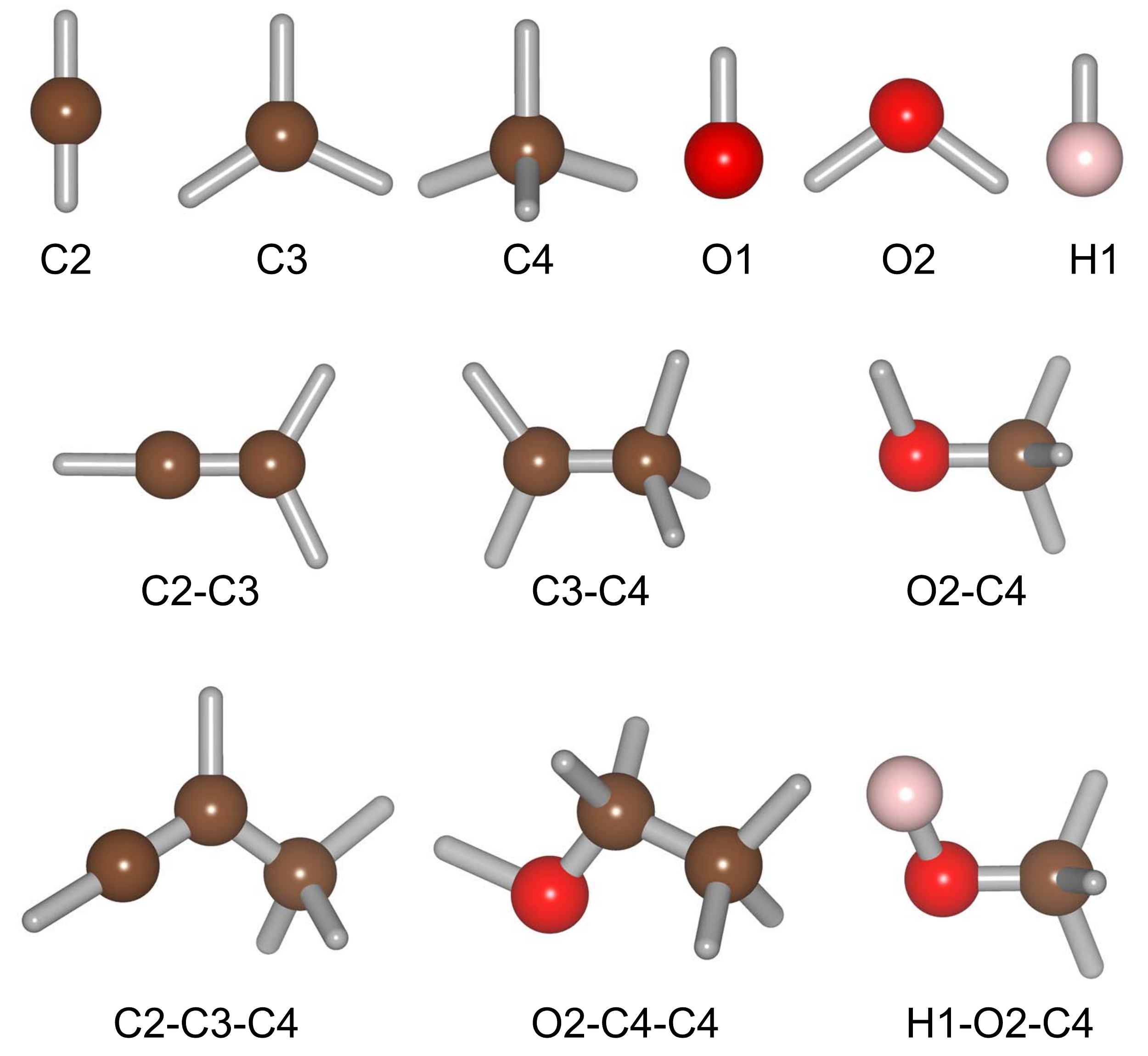}
  \caption{(Color online) Illustration of the atom types (${\cal A}i$, top row), some of the bond types (${\cal A}i\mbox{-}{\cal B}j$, middle row) and two-bond catenations (${\cal A}i\mbox{-}{\cal B}j\mbox{-}{\cal C}k$, bottom row) of materials composed by carbon, oxygen, and hydrogen.} \label{fig:fingerprint}
  \end{center}
\end{figure}

\section{Fingerprints}\label{sec:fingerprint}
A hierarchy of equilibrium structure fingerprints of the same family with increasing levels of sophistication are proposed here. The construction of fingerprints was guided by two simple chemical concepts, i.e., chemical bonds and coordination number. The former intuitively characterizes the short-range interatomic interactions \cite{Pauling} while the latter is the number of bonds involving a given atom. In major classes of materials composed of light elements like C, H, O, N, and F, these concepts are well-defined. In particular, the length of a given bond involving these elements falls in a narrow range (see Refs. \onlinecite{Allen:Table, AllenBook} for a comprehensive bond length statistics). For instance, the equilibrium length of a single bond between two C atoms is $\simeq 1.50$\AA, the length of a double bond between two C atoms is $\simeq 1.45$\AA, and the length of a double bond between a C atom and an O atom is $\simeq 1.20$\AA.\cite{Allen:Table, AllenBook} The coordination number is also well-defined, i.e., for a C atom, it can only be 2, 3, or 4 while each O atom can generally bond with 1 or 2 other atoms. Therefore, atoms in a structure can be unambiguously classified (or labeled) by ${\cal A}i$ where $\cal A$ is the type of the element (${\cal A}\in \{{\rm C, O, H}\}$) and $i$ is its coordination number. Likewise, bonds can be specified by the types of its two ends, e.g., ${\cal A}i\mbox{-} {\cal B}j$. For the datasets of C, O, and H, the six possible atom types are C2, C3, C4, O1, O2, and H1 while there are sixteen chemically permissible types of bonds, namely C2-C2, C2-C3, C2-C4, C2-O1, C2-O2, C2-H1, C3-C3, C3-C4, C3-O1, C3-O2, C3-H1, C4-C4, C4-O2, C4-H1, O2-O2, and O2-H1. Except C2-O1, C2-O2, and O2-O2, thirteen of them are present in our molecules and crystals datasets. The atom and bond types belong to a family of related structural building units (subsequently described) that can be used to numerically represent the materials structures and hence, are used to define the fingerprints. In particular, the $i^{\rm th}-$order fingerprint ${\bf f}^{(i)}$ is defined in terms of its components as
\begin{equation}\label{Eq:definition}
f^{(i)}_\kappa=\frac{n^{(i)}_\kappa}{N_{\rm at}}.
\end{equation}
Here, $n^{(i)}_\kappa$ is the number of building units (or fragments or motifs) of type $\kappa$ and $N_{\rm at}$ is the number of atoms either in the molecule or in the unit cell of a crystal. Four types of fingerprints, namely ${\bf f}^{(0)}$, ${\bf f}^{(1)}$, ${\bf f}^{(2)}$, and ${\bf f}^{(3)}$, are discussed in the following subsections.

\subsection{$0^{\rm th}$-order fingerprint, ${\bf f}^{(0)}$}
The simplest ($0^{\rm th}$-order) fingerprint ${\bf f}^{(0)}$ represents the fractions of all the element types $\cal A$ existing in the structures, i.e., $\kappa\equiv{\cal A}$. Therefore, in the definition (\ref{Eq:definition}) of ${\bf f}^{(0)}$, $n^{(0)}_{\kappa\equiv{\cal A}}$ is the number of atoms of element $\cal A$. This fingerprint is a three-dimensional vector whose components satisfy a simple normalization condition $\sum_{{\cal A}\in{\{\rm C,O,H}\}}f^{(i)}_{\cal A} = 1$.

\subsection{$1^{\rm st}$-order fingerprint, ${\bf f}^{(1)}$}
Next in the hierarchy is the case $\kappa\equiv {\cal A}i$ in which $n^{(1)}_{\kappa\equiv{\cal A}i}$ is the number of $\cal A$ atoms which are $i-$fold coordinated. ${\bf f}^{(1)}$ is a 6-dimensional vector, satisfying several constraints established from the definition or from the chemistry. The first one is the normalization condition, given as
\begin{equation}\label{eq:norm1}
\sum_{{\cal A}i} f^{(1)}_{{\cal A}i}=1.
\end{equation}
Within the two datasets, all the C2 atoms should be grouped by pairs, forming triple ${\rm C}$$\equiv$${\rm C}$ bonds. Therefore, the number of C2 atoms, which is $N_{\rm at}\times f^{(1)}_{{\rm C}2}$, must be an even integer. Moreover, since each C3 atom only make a double bond with either an O1 atom or another C3 atom, one must have $f^{(1)}_{{\rm C}3}\geq f^{(1)}_{{\rm O}1}$ while $N_{\rm at}\times \left[f^{(1)}_{{\rm C}3}-f^{(1)}_{{\rm O}1}\right]$ is an even number. By examining the connectivity of a structure, another constraint reads
\begin{equation}
\label{eq:constr2}
f^{(1)}_{{\rm H}1}-2f^{(1)}_{{\rm C}4}- f^{(1)}_{{\rm C}3}+f^{(1)}_{{\rm O}1} = \frac{2}{N_{\rm at}}\left(1 - N_\bigcirc - d\right)
\end{equation}
where $N_\bigcirc$ is the number of closed loops of bonds and $d$ is a structure-dependent parameter. For molecules and crystals composed of isolated substructures (or molecules), $d=0$ while for crystals composed of connected substructures, $d > 0$. The derivation of this constraint is given in Appendix \ref{append:constrain}. The last constraint of ${\bf f}^{(1)}$ is written in the form of a recursion relation, i.e.,
\begin{equation}
\sum_i f^{(1)}_{{\cal A}i} = f^{(0)}_{\cal A}.
\end{equation}

\subsection{$2^{\rm nd}$-order fingerprint, ${\bf f}^{(2)}$}
Both ${\bf f}^{(0)}$ and ${\bf f}^{(1)}$ are local, representing the density of the atom types of a material. The equilibrium interatomic distance is somehow captured by the $2^{\rm nd}$-order fingerprint ${\bf f}^{(2)}$ where all the possible bonds are counted.  ${\bf f}^{(2)}$ is a 13-dimensional vector whose components, $f^{(2)}_{{\cal A}i\mbox{-}{\cal B}j}$, represent the normalized number $n^{(2)}_{{\cal A}i\mbox{-}{\cal B}j}$ of the ${\cal A}i\mbox{-}{\cal B}j$ bonds in the structure. From ${\bf f}^{(2)}$, ${\bf f}^{(1)}$ can readily be determined by a recursion relation
\begin{equation}
\label{eq:f2rec}
f^{(1)}_{{\cal A}i} =  \sum_{{\cal B}j} \frac{2^{\delta_{{\cal A}i,{\cal B}j}-1}}{i} f^{(2)}_{{\cal A}i\mbox{-}{\cal B}j}
\end{equation}
where $\delta_{{\cal A}i,{\cal B}j}$ is used to remove the double counting when ${\cal A}i\equiv{\cal B}j$ [see Appendix \ref{append:recur} for the derivation of (\ref{eq:f2rec})]. Through this recursion relation, all the constraints that ${\bf f}^{(1)}$ obeys are applicable for ${\bf f}^{(2)}$. We note that ${\bf f}^{(2)}$ was discussed in several previous works, e.g., in Refs. \onlinecite{bond_energy, CoulombMatrix, Moussa:2012} under the name of ``bond counting".  This fingerprint can also be regarded as a generalization of ``doubles", the fingerprint defined in Ref. \onlinecite{Pilania_SR} for the chain models of polymers.

\subsection{$3^{\rm rd}$-order fingerprint, ${\bf f}^{(3)}$}
In the $3^{\rm rd}$-order fingerprint ${\bf f}^{(3)}$, the number of two-bond catenation is represented, i.e., $\kappa\equiv{\cal A}i\mbox{-}{\cal B}j\mbox{-}{\cal C}k$. In particular, the definition (\ref{Eq:definition}) for $f^{(3)}_{\kappa\equiv{\cal A}i\mbox{-}{\cal B}j\mbox{-}{\cal C}k}$ involves $n_{{\cal A}i\mbox{-}{\cal B}j\mbox{-}{\cal C}k}$, which is the number of ${\cal A}i\mbox{-}{\cal B}j\mbox{-}{\cal C}k$ sequences, or equivalently, the catenation of two bonds ${\cal A}i\mbox{-}{\cal B}j$ and ${\cal B}j\mbox{-}{\cal C}k$. Considering compounds of C, O, and H, there are 125 possible distinct catenation of two bonds ${\cal A}i\mbox{-}{\cal B}j$ and ${\cal B}j\mbox{-}{\cal C}k$. From ${\bf f}^{(3)}$, ${\bf f}^{(2)}$ can be determined as (see Appendix \ref{append:recur})
\begin{equation}
\begin{array}{ll}
f^{(2)}_{{\cal A}i\mbox{-}{\cal B}j} & = \displaystyle{\sum_{{\cal C}k}}\left[\frac{2^{\delta_{{\cal A}i,{\cal C}k}-1}}{j-1}f^{(3)}_{{\cal A}i\mbox{-}{\cal B}j\mbox{-}{\cal C}k}\right] \\
& =\displaystyle{\sum_{{\cal C}k}\left[\frac{2^{\delta_{{\cal B}j,{\cal C}k}-1}}{i-1}f^{(3)}_{{\cal B}j\mbox{-}{\cal A}i\mbox{-}{\cal C}k}\right].}
\end{array}
\end{equation}
Similar to ${\bf f}^{(2)}$, ${\bf f}^{(3)}$ can be viewed as a generalization of ``triples", the fingerprint examined in Ref. \onlinecite{Pilania_SR}.

\begin{figure}[t]
  \begin{center}
    \includegraphics[width= 8 cm]{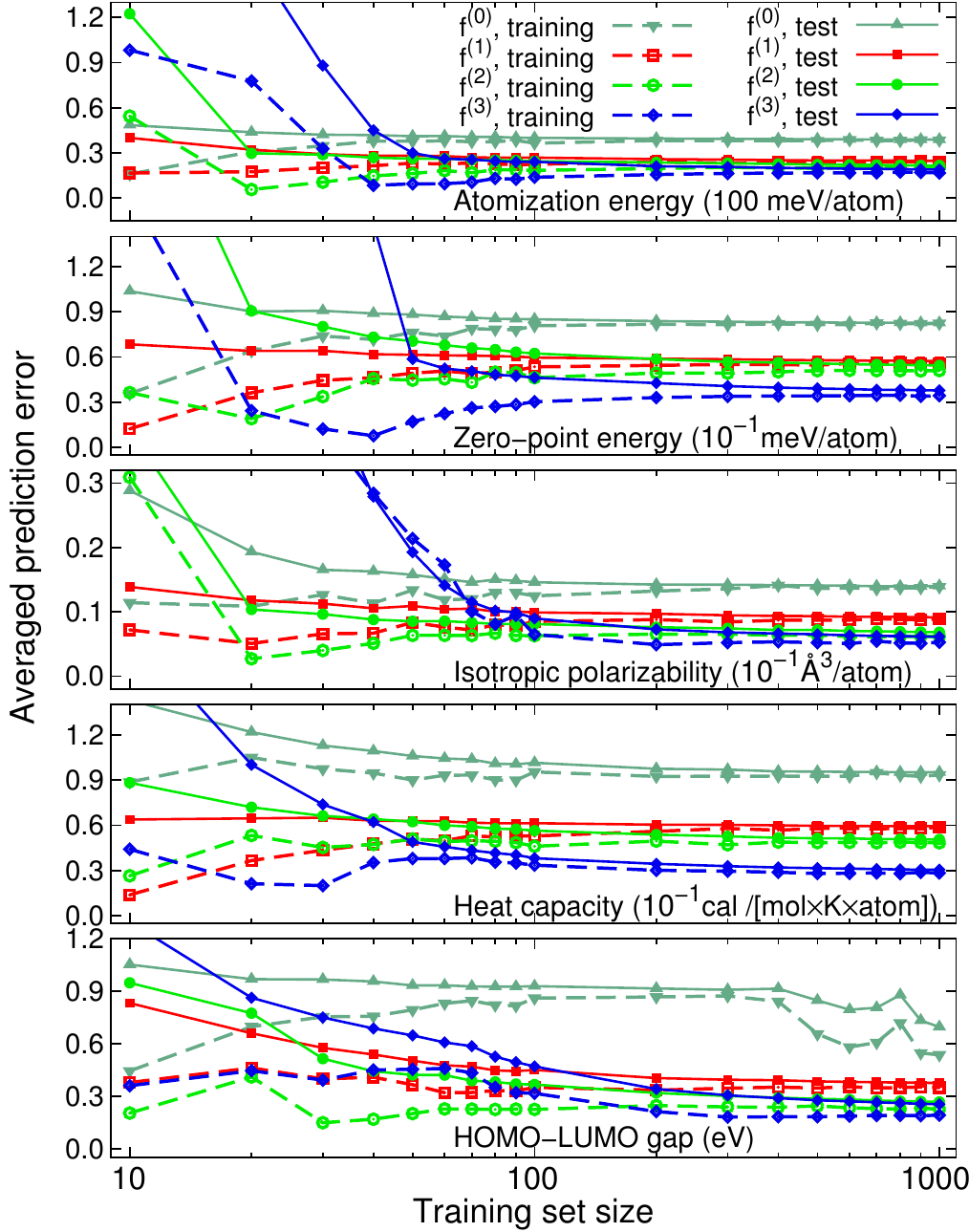}
  \caption{(Color online) Learning curves corresponding to ${\cal E}_{\rm at}$, ${\cal E}_{\rm ZP}$, $\alpha$, $C_v$, and $E_{\rm HL}$. For each model, ${\bf f}^{(0)}$, ${\bf f}^{(1)}$, ${\bf f}^{(2)}$, and ${\bf f}^{(3)}$ are used to represent the molecules.  Calculated data is given by symbols while curves are the guide for the eyes.} \label{fig:error_molec}
  \end{center}
\end{figure}

\begin{figure*}[t]
  \begin{center}
    \includegraphics[width= 16 cm]{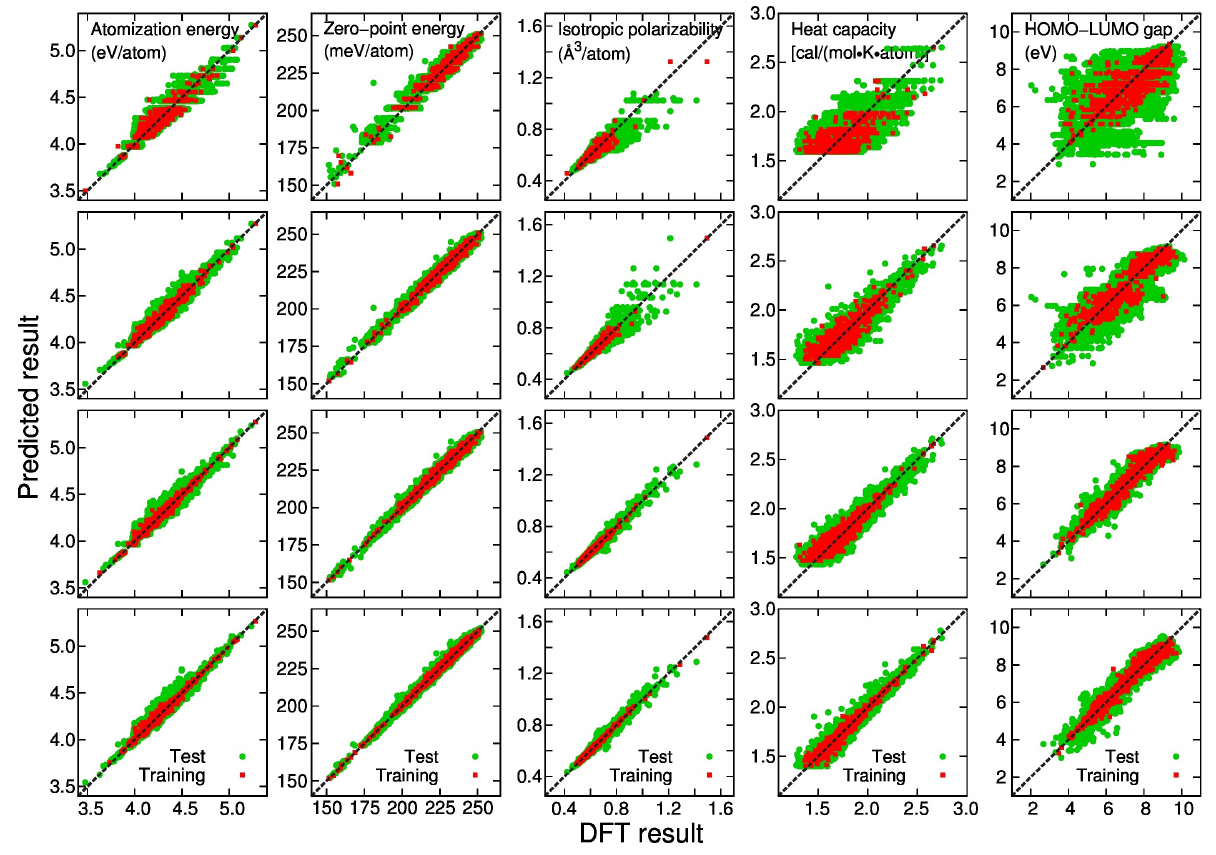}
  \caption{(Color online) Predictions for ${\cal E}_{\rm at}$, ${\cal E}_{\rm ZP}$, $\alpha$, $C_v$, and $E_{\rm HL}$  of the molecules dataset, using ${\bf f}^{(0)}$, ${\bf f}^{(1)}$, ${\bf f}^{(2)}$, and ${\bf f}^{(3)}$ (from top row to bottom row). For each prediction, the training dataset consists of 1,000 points while the test dataset includes the remaining 44,708 data points.} \label{fig:predic_molec}
  \end{center}
\end{figure*}

\section{Property prediction model}\label{sec:learningmodel}
A learning model is critical in order to map the fingerprints to properties. In this work, we chose Gaussian kernel ridge regression (KRR),\cite{hofmann2008, Muller:KRR, hastie:MLbook} the technique which has successfully been used in material properties predictions \cite{CoulombMatrix, Pilania_SR, Venke:14,Lilienfeld:delta,Lilienfeld:molec} Within this model, the input fingerprints are transformed into higher-dimensional space whereby a linear relation between the transformed fingerprints and the associated properties can be established. This mapping involves the distances between fingerprints and can be regarded as a similarity-based prediction model, i.e., similar properties may be predicted for materials with similar fingerprints.

In the KRR model, the property ${\cal P}_\mu$ of a structure $\mu$ is predicted as an weighted sum of Gaussians
\begin{equation}
\label{Eq:predict}
{\cal P}_\mu = \sum_\nu \alpha_\nu\exp\left[-\frac{1}{2}\left(\frac{d_{\mu\nu}}{\sigma}\right)^2\right],
\end{equation}
where $\nu$ runs over all the fingerprints in the training dataset. Here, $d_{\mu\nu}$ is the distance between fingerprints $\mu$ and $\nu$, defined as the Euclidean metric $d_{\mu\nu} = \sqrt{\sum_\kappa\left(f^\mu_\kappa-f^\nu_\kappa\right)^2}$. The Gaussian width parameter $\sigma$ and the regression coefficients $\alpha_\nu$ are determined within the training phase whence a regularized objective function is minimized.\cite{hofmann2008, Muller:KRR, hastie:MLbook} During this phase, $\sigma$ and the regularization parameter are determined by $k$-fold cross validation on the training set ($k=5$ in this work). Within this method, the training dataset is split into $k$ bins, any of the bins is considered to be a new test dataset while the remaining $k-1$ bins form a new training datatest. This procedure is repeated for each of the $k$ bins and for every value of $\sigma$ and $\lambda$ on a preselected logarithmic-scale grid. The optimal values of $\sigma$ and $\lambda$, i.e., those leading to the minimum $k$-fold cross-validation (mean absolute) error, are used to compute $\alpha_\nu$ of the entire dataset.

\section{Property prediction results}\label{sec:predict}

\subsection{Molecules dataset}
The four fingerprints considered, namely ${\bf f}^{(0)}$, ${\bf f}^{(1)}$, ${\bf f}^{(2)}$, and ${\bf f}^{(3)}$, were used to represent the molecules dataset. To mimic the learning and prediction processes, the dataset was randomly partitioned into a training dataset and a test dataset. The KRR model was then trained on the training dataset using five-fold cross validation before predictions were made on the test dataset. We show in Fig. \ref{fig:error_molec} the learning curves of ${\cal E}_{\rm at}$, ${\cal E}_{\rm ZP}$, $\alpha$, $C_v$, and $E_{\rm HL}$, plotting the training and test errors against the number of molecules in the training dataset (data reported in this figure was averaged over 30 independent runs). In addition, predictions for the test dataset of 44,708 molecules after training the KRR model on a dataset of 1,000 molecules are shown in Fig. \ref{fig:predic_molec}. As discussed in detail below, both Fig. 2 and Fig. 3 indicate that all of these properties can be very well predicted by using either ${\bf f}^{(2)}$ or ${\bf f}^{(3)}$, provided that the KRR model is trained on a training dataset of $\simeq 200$ or more data points.

The general tendency, as revealed by Fig. \ref{fig:error_molec}, is that higher-order fingerprints offer more accurate predictions. The $0^{\rm th}$-order fingerprint ${\bf f}^{(0)}$ can be used to roughly estimate energy-related quantities, i.e., ${\cal E}_{\rm at}$ and ${\cal E}_{\rm ZP}$ while it can not be used for others. For instance, $E_{\rm HL}$ can not be predicted with ${\bf f}^{(0)}$ because this fingerprint is totally local in nature, encoding no information at any finite range. Consequently, the finite conjugation length, known to signal the energy gap reduction in complex (conjugated) systems (see, for example Ref. \onlinecite{StallingaBook}), is not captured by ${\bf f}^{(0)}$. Fingerprints of higher orders, e.g., ${\bf f}^{(1)}$, ${\bf f}^{(2)}$ and ${\bf f}^{(3)}$, contain some information at increasing ranges, allowing for systematically better predicting $E_{\rm HL}$. These fingerprints also work sufficiently well in predicting ${\cal E}_{\rm at}$ and ${\cal E}_{\rm ZP}$. With ${\bf f}^{(1)}$, the averaged error in predicting ${\cal E}_{\rm at}$ is $\simeq 25$ meV/atom while this error is reduced to $\simeq 20$ meV/atom and $\simeq 18$ meV/atom if ${\bf f}^{(2)}$ and ${\bf f}^{(3)}$, respectively, are used. The very good power of ${\bf f}^{(2)}$ in predicting ${\cal E}_{\rm at}$ reproduces the similar conclusions drawn for the ``bond counting" fingerprint by Ref. \onlinecite{Moussa:2012}. This behavior is understandable because the dissociation energy of chemical bonds in organic molecules and crystals, which dominates the stability of these systems, are well-defined \cite{bond_energy} in the same fashion with the bond length as previously discussed. Interestingly, this predictive power can significantly be improved if more advanced fingerprints, i.e., those can capture the small perturbations of interatomic distances like Coulomb matrix, are used.\cite{Lilienfeld:delta,Lilienfeld:molec} Compared to ${\bf f}^{(1)}$ and ${\bf f}^{(2)}$, ${\bf f}^{(3)}$ is significantly better in predicting $C_v$. The considerable improvement in the predictions of $\alpha$ when ${\bf f}^{(2)}$ is used instead of ${\bf f}^{(1)}$ may indicate the key contribution from polar bonds to the high-value regime of $\alpha$.

\begin{figure}[t]
  \begin{center}
    \includegraphics[width= 8 cm]{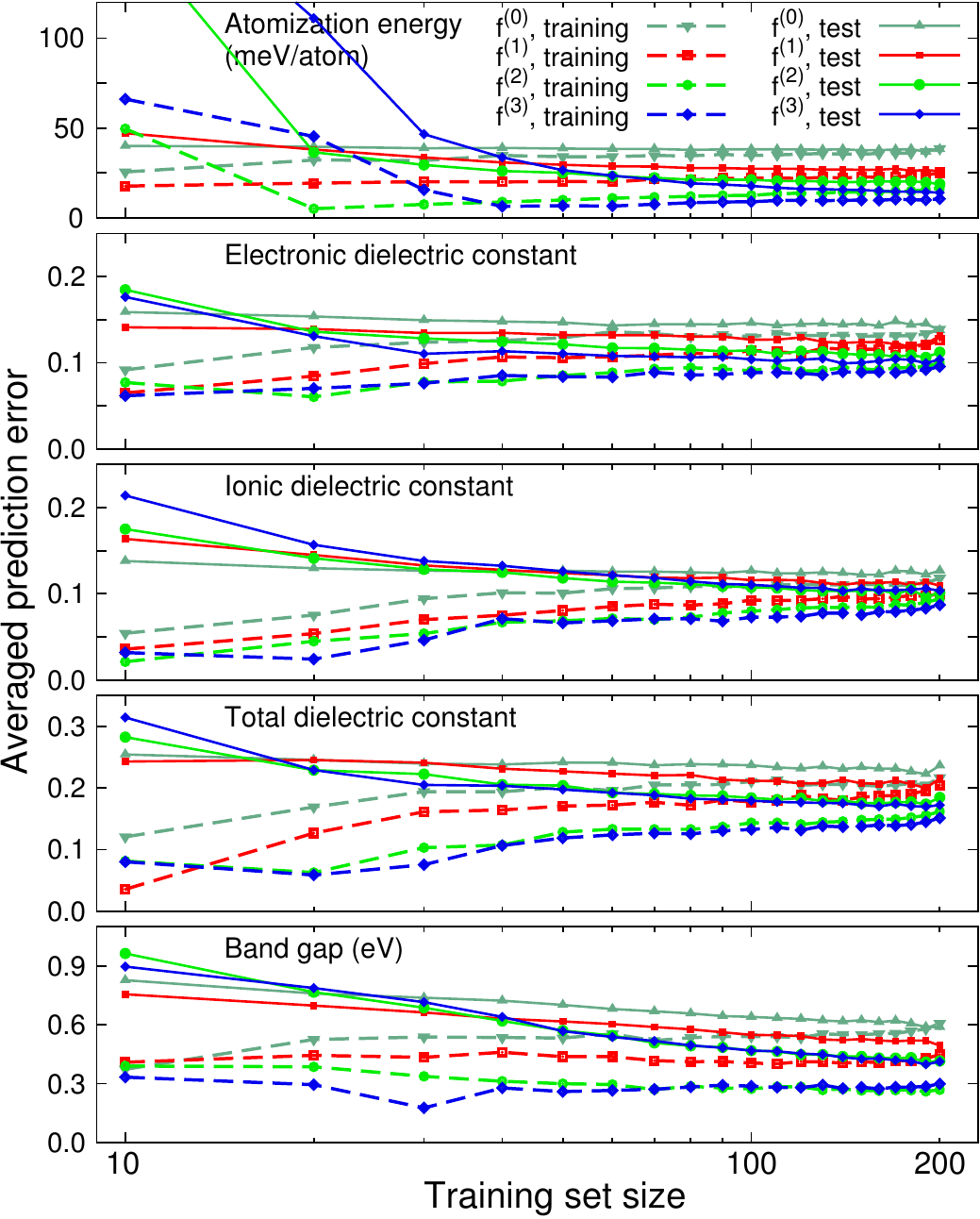}
  \caption{(Color online) Learning curves corresponding to ${\cal E}_{\rm at}$, $\epsilon_{\rm elec}$, $\epsilon_{\rm ion}$, $\epsilon$, and $E_{\rm g}$ determined by using ${\bf f}^{(0)}$, ${\bf f}^{(1)}$, ${\bf f}^{(2)}$, and ${\bf f}^{(3)}$ for representing the crystals structures.  Calculated data is shown by symbols while curves are the guide for the eyes.} \label{fig:error_crystal}
  \end{center}
\end{figure}

\begin{figure*}[t]
  \begin{center}
    \includegraphics[width= 16 cm]{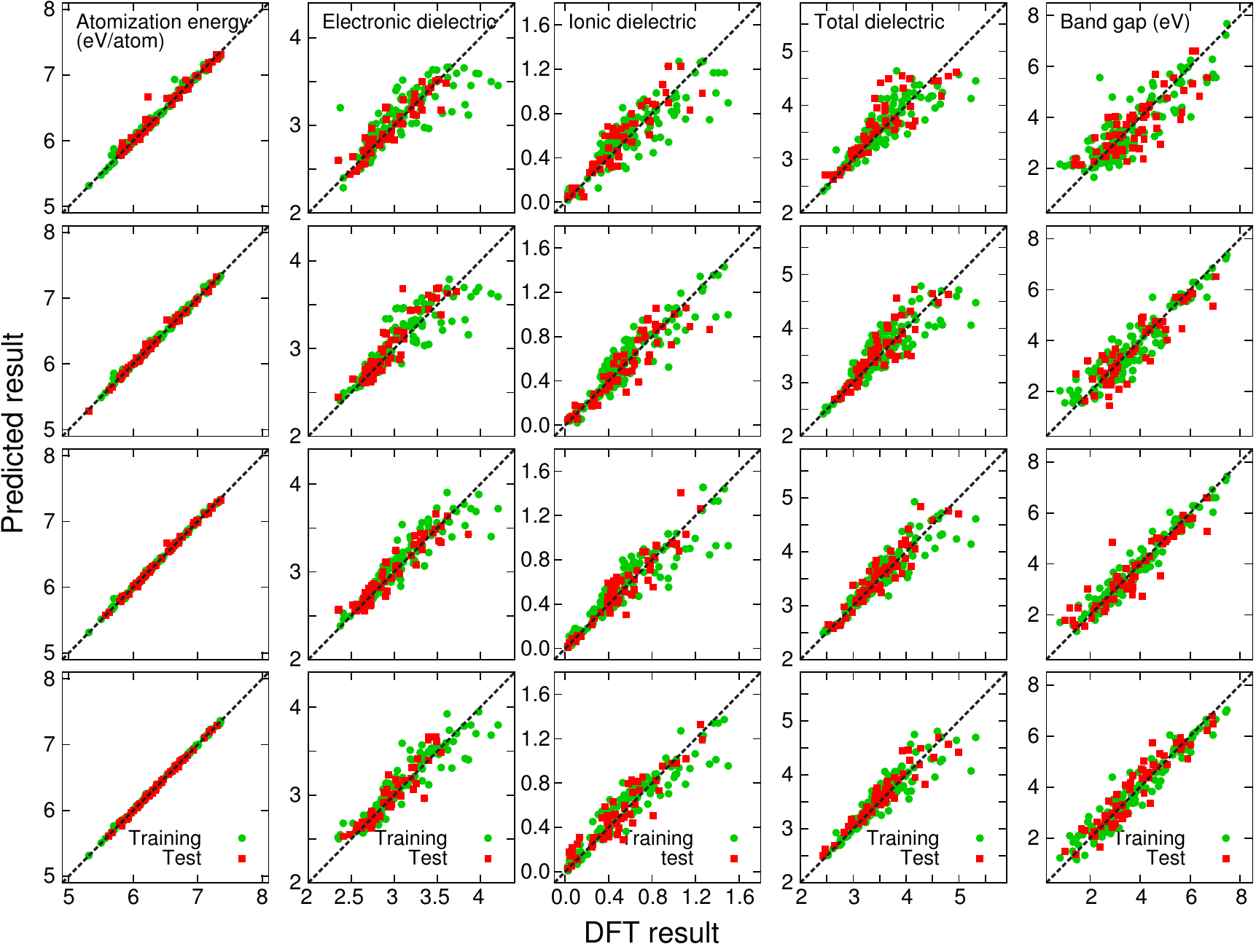}
  \caption{(Color online) Predictions for ${\cal E}_{\rm at}$, $\epsilon_{\rm elec}$, $\epsilon_{\rm ion}$, $\epsilon$, and $E_{\rm gap}$ of the crystals dataset, using ${\bf f}^{(0)}$, ${\bf f}^{(1)}$, ${\bf f}^{(2)}$, and ${\bf f}^{(3)}$ (from top row to bottom row). For each prediction, the training set size is 150 and the remaining 70 points form the test set.} \label{fig:predic_solids}
  \end{center}
\end{figure*}

\begin{figure}[b]
\begin{center}
  \includegraphics[width=8cm]{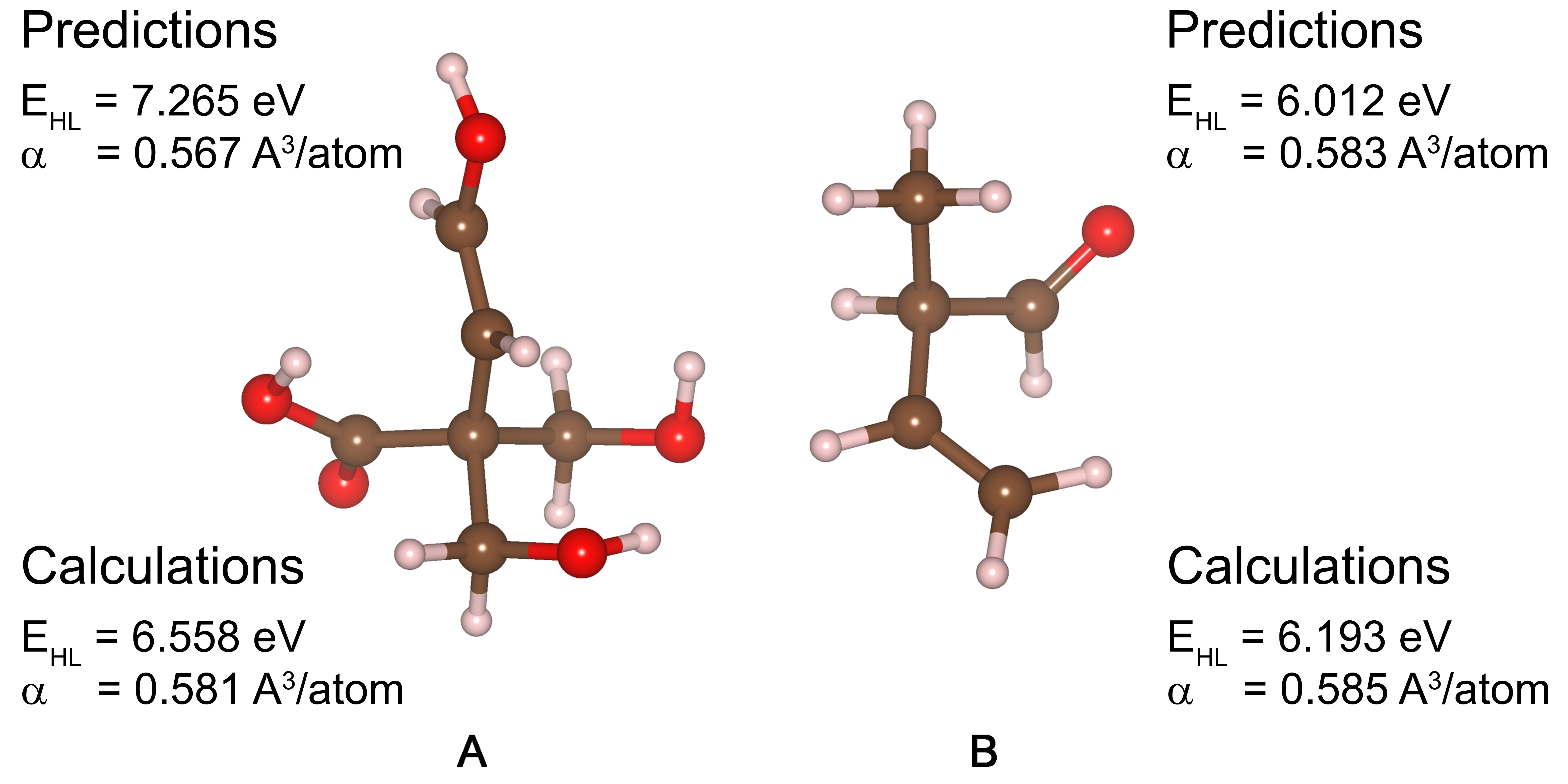}
  \caption{(Color online) Optimized molecules, constructed from two predicted fingerprints A and B, shown with the predicted and calculated values of $E_{\rm HL}$ and $\alpha$. Carbon, oxygen, and hydrogen atoms are given in dark brown, red, and pink.}\label{fig:enum}
\end{center}
\end{figure}

\subsection{Crystals dataset}
We performed similar predictions for the dataset of 215 crystals containing C, O, and H. Using the KRR model coupled with ${\bf f}^{(0)}$, ${\bf f}^{(1)}$, ${\bf f}^{(2)}$, and ${\bf f}^{(3)}$, five properties of these crystals, including the atomization energies ${\cal E}_{\rm at}$, the band gap $E_{\rm g}$, the electronic dielectric constant $\epsilon_{\rm elec}$, the ionic dielectric constant $\epsilon_{\rm ion}$, and the total dielectric constant $\epsilon_{\rm tot} = \epsilon_{\rm elec} + \epsilon_{\rm ion}$, were predicted. We show in Fig. \ref{fig:error_crystal} the learning curves, representing the errors of the predictions using these fingerprints, averaged over $100$ independent runs. In Fig. \ref{fig:predic_solids}, the predictions for the five properties are given, using the KRR model trained on a random training set of $150$ data points.

Clearly, the tendency of the prediction performances on the crystals dataset is similar to those of the molecules dataset, i.e., high accuracies are obtained with fingerprints of higher orders, and properties which are governed by long-ranged information, e.g., band gap $E_{\rm g}$, can only be predicted with high-order fingerprints. For the atomization energy ${\cal E}_{\rm at}$, predictions with ${\bf f}^{(0)}$ and ${\bf f}^{(1)}$ leads to quite high averaged errors, which reduced to $\simeq 18$ meV/atom and $\simeq 15$ meV/atom when ${\bf f}^{(2)}$ and ${\bf f}^{(3)}$, respectively, were used. Overall, all the five examined properties can be predicted well when high-order fingerprints are used to represent the crystals. For instance, by employing ${\bf f}^{(3)}$, the averaged error in predicting $E_{\rm g}$ is $\simeq 0.45$ eV while the electronic dielectric constant $\epsilon_{\rm elec}$ and the ionic dielectric constant $\epsilon_{\rm ion}$ can be predicted with an averaged error of $0.1-0.2$.

\section{Utilities of the fingerprints}\label{sec:design}
The demonstrated predictive power of the KRR model, which uses ${\bf f}^{(i)}$ to represent materials structures, inspires the idea of using this model to rationally optimize materials for a targeted property ${\cal P}_{\rm opt}$, the concept often referred to as ``inverse design". \cite{Ceder, Besenbacher20031998, FrancesNature, QUA:QUA24687} In fact, a large number of success stories along this direction have been reported in the past, using various approaches, e.g., iteratively optimizing the properties of a given compound or {\it on-the-fly} screening when searching for stable structures. \cite{Lilienfeld:inverse, Marcon:07, Lilienfeld:energy_gradient, Lilienfeld:alchemi, Wang:design_molec, Keinan_design_molec, Keinan_design_molec_2, Rinderspacher_design_molec, Curtarolo2003, Greeley, Avezac2012, XiangInvDesign, Phillips:ML} Here, our idea is that starting from a trained KRR model, fingerprints which correspond to the desired properties can be predicted. Then, molecular structures will be reconstructed from the predicted fingerprints. Finally, the targeted properties will be verified by DFT calculations at the same level with those used for the training dataset. 

The greatest challenge of this procedure is to ensure that the predicted fingerprint is physically and chemically meaningful, i.e., at least one material structure can be reconstructed from it.\cite{Lilienfeld:CCS, Lilienfeld:ensemble} Therefore, one must mathematically define the subspace of the meaningful fingerprints, and then limit the search for desired fingerprints within this subspace. We present two approaches which can be used for designing molecules (the work of designing crystals is not considered here).

\subsection{Design via enumeration}
The central idea of this approach is that the components of a given fingerprint can be enumerated in a given way so that it is meaningful. We used ${\bf f}^{(2)}$ for a demonstration because predictions using this fingerprint are good while its dimensionality is not too high like ${\bf f}^{(3)}$. We first implemented the applicable rules involving bonds and coordination numbers by defining five ``backbone" blocks. They include C4, C$=$C (a pair of C3 atoms with a double bond), C$\equiv$C (a pair of C2 atoms with a triple bond), C$=$O (one C3 and one O1 atom linked by a double bond), and O2. By definition, all of the dangling bonds starting from these blocks are single, thus any of them can be connected to others without any constraint. Then, given a set of backbone blocks, all the possible arrangements can be scanned, keeping track of the connectivity to eliminate some dangling bonds, and saturating the remaining dangling bonds by either H1 or OH, referred to as ``ending" blocks. From the obtained arrangements, ${\bf f}^{(2)}$ can be unambiguously determined and their properties were predicted. Those with targeted properties were singled out to rebuild molecular structures for validating calculations. We show in Fig. \ref{fig:enum} two optimized molecules constructed from two of the predicted fingerprints, labeled by A and B, accompanied by the predicted and calculated $E_{\rm HL}$ and $\alpha$. The results given in Fig. \ref{fig:enum} indicate that the desired molecules are indeed obtained.

\subsection{Design via inversion}
Different from the enumeration approach, this procedure aims to directly determine the fingerprints, starting from desired properties. This goal can be achieved by optimizing an objective function, aiming towards the desired properties while applying the constraints that ensure the fingerprints considered are meaningful. Because the reconstruction step requires a simple enough fingerprint, ${\bf f}^{(1)}$ was selected for this approach. Among the constraints established for ${\bf f}^{(1)}$, (\ref{eq:norm1}) and (\ref{eq:constr2}) are explicitly imposed in the objective function defined below
\begin{equation}
\label{eq:objfnc}
\begin{array}{rl}
G[{\bf f}^{(1)}, \lambda_1, \lambda_2]          & \displaystyle{=  \left( {\cal P}-{\cal P}_{\rm opt}\right)^2 + \lambda_1\left[\sum_{{\cal A}i} f^{(1)}_{{\cal A}i}-1\right]^2 }\\
           & + \lambda_2\left[f^{(1)}_{{\rm H}1}-2f^{(1)}_{{\rm C}4}- f^{(1)}_{{\rm C}3}+f^{(1)}_{{\rm O}1}\right]^2.
\end{array}
\end{equation}
Here, $\lambda_1$, and $\lambda_2$ are the Lagrange multipliers associated with the constraints while $\cal P$ is the property (or properties) of the trial fingerprint ${\bf f}^{(1)}$ predicted by the trained KRR model. In practice, we evaluated $\cal P$ by averaging many predictions, each of them was given by the KRR model trained on a randomly selected training dataset of 1,000 data points. All the terms in (\ref{eq:objfnc}) are given in the quadratic form to smoothen $G$. Generally, the problem of minimizing  $G[{\bf f}^{(1)}, \lambda_1, \lambda_2]$ (performed with simulated annealing\cite{Kirkpatrick13051983} in this work) returns many solutions ${\bf F}^{(1)}$. For each of them, $N_{\rm at}$ was determined by minimizing another objective function $D[{\bf F}]$ defined as
\begin{equation}
D[{\bf F}^{(1)}] = \sum_{{\cal A}i}\left[N_{\rm at}F^{(1)}_{{\cal A}i}-{\rm nint}\left(N_{\rm at}F^{(1)}_{{\cal A}i}\right) \right]^2,
\end{equation}
where ${\rm nint}(x)$ returns the closest integer to $x$. Once $N_{\rm at}$ is determined, a post-screening step is performed to consider the possibility of $N_\bigcirc > 0$ and to single out the fingerprints so that $N_{\rm at}F_{\rm C2}^{(1)}$ and $N_{\rm at}\left[F_{\rm C3}^{(1)}-F_{\rm O1}^{(1)}\right]$ are positive even numbers. Such fingerprints are meaningful, i.e., molecules can be built up from any of them.

\begin{figure}[t]
\begin{center}
    \includegraphics[width= 8 cm]{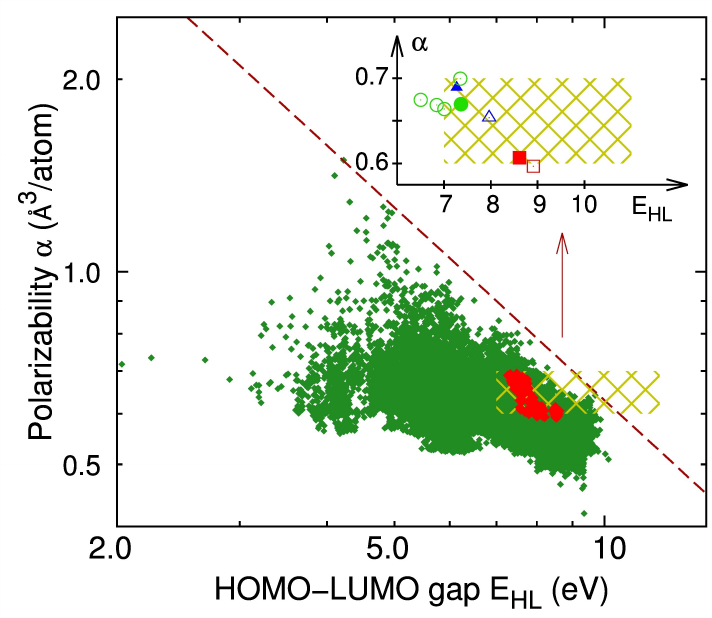}
    \caption{(color online) $E_{\rm HL}-\alpha$ log-log plot of the molecules dataset, shown by forest-green symbols while the predicted fingerprints are shown by red diamonds within the regime of desired properties, i.e., $0.6 \leq \alpha \leq 0.7$\AA$^3$/atom and $E_{\rm HL} \geq 7.0$ eV. In the inset, the predicted and calculated properties of the molecules reconstructed from three predicted fingerprints, i.e., C, D, and E, are shown by closed and open symbols: triangles for C, circles for D, and squares for E. The dashed line sketches the limit $\alpha \sim 1/E_{\rm HL}$ addressed in the text.}\label{fig:predict2}
\end{center}
\end{figure}

We demonstrate this procedure by optimizing two properties simultaneously, i.e., $E_{\rm HL}$ and $\alpha$. We note that these properties seem to be competing, as shown in Fig. \ref{fig:predict2} where an asymptotic limit of the form $\alpha \sim 1/E_{\rm HL}$ can be seen (similar limit between two related properties of crystals, namely $\epsilon_{\rm elec}$ and $E_{\rm g}$ was documented earlier in Ref. \onlinecite{Chenchen:polymer}). An examination of Fig. \ref{fig:predic_molec} reveals that the prediction of $\alpha$ using ${\bf f}^{(1)}$ is fairly good in the region of $\alpha < 0.8$ \AA$^3$/atom. For this reason, we searched for new molecules, i.e., those that do not exist in the molecules dataset, of which $0.6 \leq \alpha \leq 0.7$ \AA$^3$/atom while $E_{\rm HL} \geq 7$ eV and show the results in Fig. \ref{fig:predict2}. While the calculated $E_{\rm HL}$ of the molecules dataset can reach the upper limit of $\simeq 10$ eV, all the predictions for $E_{\rm HL}$ by the KRR model are below 9 eV. The reason is given in Fig. \ref{fig:predic_molec} which clearly implies that when ${\bf f}^{(1)}$ is coupled with the KRR model, high values of $E_{\rm HL}$ ($8 \leq E_{\rm HL} \leq 10$ eV) are generally underestimated by roughly 1 eV. Three of the predicted fingerprints, labeled by C, D, and E, were selected for rebuilding new molecules. From either C or E, only one molecule can be constructed while many different molecules correspond to D. All of the molecules reconstructed from C, D, and E were optimized and then their $\alpha$ and $E_{\rm HL}$ were calculated with Gaussian 09,\cite{GAUSSIAN09} using the 6-31G(2df,p) basis set and the B3LYP XC functional.\cite{B3LYP1,B3LYP2} The results are summarized in Table \ref{table:validation2} and in the inset of Fig. \ref{fig:predict2}, demonstrating that the molecules with desired values of $\alpha$ and $E_{\rm HL}$ were actually obtained. Detailed information on all of the designed molecules can be found in the Supplemental Material.\cite{supplement}

\subsection{Remarks}
It is worth noting that the key feature of ${\bf f}^{(i)}$ which is useable for the described enumeration and inversion design procedures is their discontinuity with respect to slight configurational perturbations. Because all the possible chemical bonds appearing in a molecule comprising C, O, and H are well-defined, it is very likely that the optimization step performed on the reconstructed molecules preserves the predicted fingerprint. Moreover, the efficiency of the designing approaches depends on several factors, including the prediction accuracy of the fingerprints used. Although predictions by using high-order fingerprints are systematically better, the complexity generated by their high dimensionality is significant. Comparing to the procedure described above, that utilizing ${\bf f}^{(2)}$ or ${\bf f}^{(3)}$ needs roughly 10 and 100 more constraints for ensuring the considered fingerprints are meaningful. If the dimensionality of ${\bf f}^{(2)}$ can considerably be reduced, it may then be used for the inversion approach.

\section{Conclusions}
To summarize, we have systematically studied a family of motif-based topological fingerprints which can numerically represent major classes of molecules and crystals. By using a similarity based learning algorithm, these fingerprints can be mapped onto various properties of molecules and crystals, significantly accelerating their properties prediction. A major advantage of these fingerprints is clearly demonstrated via two procedures for designing molecules, one by enumeration and the other by inversion. These procedures rely on the accelerated properties prediction to identify the desired fingerprints, and then to reconstruct molecules that possess one or more targeted properties. We note that although only molecules and crystals comprising C, O, and H are considered in this contribution, our results can straightforwardly be generalized to those containing other light elements whose coordination preferences are well established, e.g., N and F.

\begin{table}[t]
\caption{Predicted and calculated values of $\alpha$ (in \AA$^3$/atom) and $E_{\rm HL}$ (in eV) of the molecules designed from three predicted fingerprints C, D, and E. Data from this Table is also shown in the inset of Fig. \ref{fig:predict2}.}\label{table:validation2}
\begin{tabular}{ccccccc}
\hline
\hline
Label & $N_{\rm at}$ & \multicolumn{2}{c}{Predicted} & & \multicolumn{2}{c}{Calculated} \\
\cline{3-4}\cline{6-7}
      &              &$\alpha$ &$E_{\rm HL}$  & & $\alpha$& $E_{\rm HL}$ \\
\hline
C     &   11         & $0.689$     & $7.273$ && $0.654$ & $7.964$\\
D     &   18         & $0.670$     & $7.363$ && $0.664-0.699$ & $6.502-7.348$\\
E     &   14         & $0.607$     & $8.612$ && $0.597$ & $8.909$\\
\hline
\hline
\end{tabular}
\end{table}

\begin{acknowledgements}
The authors thank Venkatesh Botu, Ghanshyam Pilania, and Vinit Sharma for useful discussions and O. Anatole von Lilienfeld for drawing our attention to some important relevant works. The present work was supported by a Multi-University Research Initiative (MURI) grant from the Office of Naval Research, under award number N00014100944. Part of the computational work was done with our sponsored TeraGrid XSEDE allocation.\cite{xsede_allocation}
\end{acknowledgements}

\appendix
\section{Constraint of ${\bf f}^{(1)}$ derived from elementary chemical rules}\label{append:constrain}
Constraint (\ref{eq:constr2}) was derived with an assumption that the desired molecular structure is connected, i.e., any pair of atoms are connected by at least one sequence of the allowed chemical bonds. Let us take a molecule in which $n_{{\cal A}i}$ is the number of the blocks ${\cal A}i$. Starting from the applicable chemical rules, all the two-fold coordinated carbon atoms are grouped by pairs, forming $n_{\rm C2}/2$ units of ${\rm C}\equiv{\rm C}$, each of which is a pair of carbon atoms linked by a triple bond. Next, $n_{\rm O1}$ one-fold coordinated oxygen atoms must bond with $n_{\rm O1}$ three-fold coordinated carbon atoms to form $n_{\rm O1}$ units of ${\rm C=O}$. Then, the remaining $n_{\rm C3}-n_{\rm O1}$ three-fold coordinated carbon atoms are grouped together by pairs, forming $(n_{\rm C3}-n_{\rm O1})/2$ units of ${\rm C}={\rm C}$. Therefore, the set of the blocks ${\cal A}i$ now contains $n_{\rm C2}/2 + n_{\rm O1} + (n_{\rm C3}-n_{\rm O1})/2 +n_{\rm C4}+ n_{\rm O2}$ units of ${\rm C}\equiv{\rm C}$, ${\rm CO}$, ${\rm C}={\rm C}$, $\rm C4$ and ${\rm O2}$. Assuming that these units are isolated, the total number of dangling bonds starting from them is
$2(n_{\rm C2}/2) + 2n_{\rm O1} + 4[(n_{\rm C3}-n_{\rm O1})/2] +4n_{\rm C4} + 2n_{\rm O2}$, or simply
\begin{equation}
n_{\rm C2} + 2n_{\rm C3} +4n_{\rm C4} + 2n_{\rm O2}.
\end{equation}
By joining $n_{\rm C2}/2 + n_{\rm O1} + (n_{\rm C3}-n_{\rm O1})/2 +n_{\rm C4}+ n_{\rm O2}$ units together, the number of dangling bonds that will be annihilated to form inter-unit bonds is $2[n_{\rm C2}/2 + n_{\rm O1} + (n_{\rm C3}-n_{\rm O1})/2 +n_{\rm C4}+ n_{\rm O2}-1]+2n_\bigcirc$ where $n_\bigcirc$ is the number of loops of bonds, each of which costs extra 2 bonds. Therefore, the number of remaining dangling bonds is
\begin{equation}
n_{\rm C3} + 2n_{\rm C4} -n_{\rm O1} -2n_\bigcirc + 2.
\end{equation}
All of these dangling bonds must be saturated by $n_{\rm H1}$ hydrogen atoms, thus
\begin{equation}\label{Eq:constrain4}
n_{\rm H1} = n_{\rm C3} + 2n_{\rm C4} -n_{\rm O1} -2n_\bigcirc + 2.
\end{equation}
The constraint (\ref{eq:constr2}) can then be obtained when we divide Eq. (\ref{Eq:constrain4}) by $N_{\rm at}$. This constraint is applicable not only for molecules but also for crystals formed by repeatedly placing an isolated molecule in a periodic grid. If these molecules are not isolated, i.e., they form a network of $d$ dimensions, $2d$ dangling bonds are used to form the network (assuming that the network are formed only by single bonds). Thus, Eq. \ref{Eq:constrain4} is given as
\begin{equation}\label{Eq:constrain5}
n_{\rm H1} = n_{\rm C3} + 2n_{\rm C4} -n_{\rm O1} -2n_\bigcirc - 2d + 2.
\end{equation}
In the general case when not only single bonds involve the network formation, the parameter $d$ used in Eq. \ref{Eq:constrain5} is not necessarily an integer.

\section{Derivation of the recursion relations of ${\bf f}^{(2)}$ and ${\bf f}^{(3)}$}\label{append:recur}
\subsection{Recursion relations of ${\bf f}^{(2)}$}
The number $n_{{\cal A}i}$ of blocks ${\cal A}i$ can be determined by counting all the bonds of ${\cal A}i\mbox{-}{\cal B}j$ type. By summing all the number of ${\cal A}i\mbox{-}{\cal B}j$ bonds, the ${\cal A}i\mbox{-}{\cal A}i$ bonds are counted twice. Therefore
\begin{equation}\label{eq:appd1}
n_{{\cal A}i} = \frac{1}{i}\left[\sum_{{\cal B}j}n_{{\cal A}i\mbox{-}{\cal B}j} -\frac{1}{2}n_{{\cal A}i\mbox{-}{\cal A}i}\right].
\end{equation}
Then, the recursion relation of ${\bf f}^{(2)}$ can be obtained by dividing (\ref{eq:appd1}) by the total number of atoms $N_{\rm at}$.
\subsection{Recursion relations of ${\bf f}^{(3)}$}
Similar to the derivation of (\ref{eq:appd1}), the fingerprint component $f^{(2)}_{{\cal A}i\mbox{-}{\cal B}j}$ can be determined by counting the number of ${{\cal A}i\mbox{-}{\cal B}j\mbox{-}{\cal C}k}$ sequences before dividing by $j-1$. In such a procedure, the ${{\cal A}i\mbox{-}{\cal B}j\mbox{-}{\cal A}i}$ sequences are counted twice. Thus, after removing the double counting, we obtain
\begin{equation}\label{eq:appd2}
n_{{\cal A}i\mbox{-}{\cal B}j} = \frac{1}{j-1}\left[\sum_{{\cal C}k}n_{{\cal A}i\mbox{-}{\cal B}j\mbox{-}{\cal C}k}-\frac{1}{2}n_{{\cal A}i\mbox{-}{\cal B}j\mbox{-}{\cal A}i}\right].
\end{equation}
We note that one can also count the number of ${{\cal B}j\mbox{-}{\cal A}i\mbox{-}{\cal C}k}$ sequences before dividing the total number by $i-1$. Thus
\begin{equation}\label{eq:appd3}
n_{{\cal A}i\mbox{-}{\cal B}j} = \frac{1}{i-1}\left[\sum_{{\cal C}k}n_{{\cal B}j\mbox{-}{\cal A}i\mbox{-}{\cal C}k}-\frac{1}{2}n_{{\cal B}j\mbox{-}{\cal A}i\mbox{-}{\cal B}j}\right].
\end{equation}
By dividing (\ref{eq:appd2}) and (\ref{eq:appd3}) by $N_{\rm at}$, two equivalent recursion relations are obtained. Moreover, we note that (\ref{eq:appd2}) and (\ref{eq:appd3}) set up a constraint that ${\bf f}^{(3)}$ must also satisfy.


\end{document}